\newcommand{\qlibname}{Qlib }
\newcommand{\qlibnamenb}{Qlib}  

\documentclass{article}
\usepackage{ijcai20}

\usepackage{times}
\usepackage{soul}
\usepackage{url}
\usepackage{hyperref}
\hypersetup{hidelinks}
\usepackage[utf8]{inputenc}
\usepackage[small]{caption}
\usepackage{graphicx}
\usepackage{amsmath}
\usepackage{amsthm}
\usepackage{booktabs}
\usepackage{algorithm}
\usepackage{algorithmic}
\usepackage{tabularx}
\usepackage{footnote}
\makesavenoteenv{tabularx}
\usepackage{diagbox}
\usepackage{amssymb}
\usepackage{float}

\title{\qlibname: An AI-oriented Quantitative Investment Platform}

\author{
Xiao Yang \and
Weiqing Liu \and
Dong Zhou \and
Jiang Bian \And
Tie-Yan Liu
\affiliations
Microsoft Research\\
\emails
\{Xiao.Yang, Weiqing.Liu, Zhou.Dong, Jiang.Bian, Tie-Yan.Liu\}@microsoft.com
}

\begin{document}

\maketitle

\begin{abstract}
Quantitative investment aims to maximize the return and minimize the risk in a sequential trading period over a set of financial instruments.
Recently, inspired by rapid development and great potential of AI technologies in generating remarkable innovation in quantitative investment, there has been increasing adoption of AI-driven workflow for quantitative research and practical investment. 
In the meantime of enriching the quantitative investment methodology, AI technologies have raised new challenges to the quantitative investment system. Particularly, the new learning paradigms for quantitative investment  
call for an infrastructure upgrade to accommodate the renovated workflow; moreover, the data-driven nature of AI technologies indeed indicates a requirement of the infrastructure with more powerful performance;
additionally, 
there exist some unique challenges for applying AI technologies to solve different tasks in the financial scenarios. 
To address these challenges and bridge the gap between AI technologies and quantitative investment, we design and develop \qlibname that aims to realize the potential, empower the research, and create the value of AI technologies in quantitative investment.  


\end{abstract}



\section{Introduction}
Quantitative investment, one of the hottest research fields, has been attracting numerous brilliant minds from both the academia and financial industry. In the last decades, with continuous efforts in optimizing the quantitative methodology, the whole community of professional investors has summarized a well-established yet imperfect quantitative research workflow.
Recently, emerging AI technologies start a new trend in this research field. With increasing attention to exploring AI's great potential in quantitative investment, AI technologies have been widely adopted in the practical investment by quantitative researchers. 

While AI technologies have been enriching the quantitative investment methodology, they also put forward new challenges to the quantitative investment system from multiple perspectives. First, the technological revolution in the quantitative investment workflow, caused by the flexibility of AI technologies, tends to require new supportive infrastructure. 
For example, while the traditional quantitative investment usually splits the whole workflow into a couple of sub-tasks, including stock trend prediction, portfolio optimization, etc., AI technologies make it possible to establish an end-to-end solution that generates the final portfolio directly. To support such end-to-end solution, it is necessary to upgrade the current infrastructure due to its data-driven nature.

Meanwhile, the AI technologies have to deal with the unique problems in some new scenarios, which require both plenty of domain knowledge in finance and rich experience in data science.  Applying the solutions to quantitative research tasks without any domain adaptation rarely works.
Such a circumstance leads to urgent demands for a platform to accommodate such a modern quantitative research workflow in the age of AI and provide guidance for the application of AI technologies in the financial scenario.

Therefore,  we propose a new AI-oriented Quantitative Investment Platform called \qlibnamenb\footnote{The code is available at https://github.com/microsoft/qlib}. 
It aims to assist the research efforts of exploring the great potential of AI technologies in quantitative investment as well as empower quantitative researchers to create more significant values on AI-driven quantitative investment.
Specifically, the AI-oriented framework of \qlibname is designed to accommodate the AI-based solutions. Moreover, it provides high-performance infrastructure dedicated for quantitative investment scenario, which makes many AI research topics possible. In addition, a batch of tools designed for machine learning in the quantitative investment scenario is integrated with \qlibname to benefit users in making fully utilization of AI technologies.

At last, we demonstrate some use cases and evaluate the performance of the infrastructure of \qlibname by comparing several solutions for a typical task in quantitative investment. The results show the infrastructure of \qlibname dedicated to quantitative investment outperforms most of existing solutions on this task. 

\section{Background and Related Works}
\label{background} 

In this section, we will first demonstrate the major practical problems of a modern quantitative researcher when applying of AI technologies in quantitative investment, which motivates the birth of \qlibnamenb. After that, we will briefly introduce the related work.

\subsection{Practical Problems}



\subsubsection{Quantitative research workflow revolution}

In the traditional investment research workflow, researchers often develop trading signals by linear models \cite{petkova2006fama} or manually designed rules\cite{murphy1999technical} based on several factors(factors are similar to features in machine learning) and basic financial data. And then, a trading strategy(typically Barra\cite{sheikh1996barra}) is followed to generate the target portfolio. At last, researchers evaluate the trading signal and portfolio by a back-testing function.

With the rise of AI technologies, it launches a technological revolution of traditional quantitative investment. The traditional quantitative research workflow is too primitive to accommodate such flexible technologies. In order to show the difference more intuitively, we'll demonstrate a typical modern research workflow based on AI technologies. It starts with a dataset with lots of features(typically more than hundreds of dimensions). Manually designing such amount of features takes lots of time. It is common to leverage machine learning algorithms to generate such features automatically\cite{potvin2004generating,neely1997technical,allen1999using,kakushadze2016101}.  Generating data \cite{feng2019enhancing} is another option for constructing a dataset. Based on diverse datasets, researchers have provided hundreds of machine learning methods to mine trading signals \cite{sezer2019financial}. Researchers could generate the target portfolio based on such trading signals. But such a workflow is not the only choice. Instead of dividing a task into several stages,  RL(reinforcement learning) provides an end-to-end solution from the data to the final trading actions directly \cite{deng2016deep}. RL optimizes the trading strategy by interacting with the environment, which is a trading simulator in the financial scenario. RL needs a responsive simulator instead of a back-testing function in the traditional research workflow.  Moreover, most of the AI algorithms have complicated hyperparameters, which need to be tuned carefully.

AI technologies are so flexible and already beyond the scope of existing tools designed for traditional methodologies. Building a research workflow based on AI technologies from scratch takes much time.

\subsubsection{High performance requirements for infrastructure}
With the emerging of AI technologies, the requirements for infrastructure have changed. Such a data-driven method could leverage a huge amount of data. The amount of data could reach the order of TB magnitude in the scenario of high-frequency trading. Besides, it is very common to derive thousands of new features (e.g., Alpha101 \cite{kakushadze2016101}) from the basic price and volume data, which consist of only five dimensions in total. Some researchers even try to create new factors or features by searching expressions \cite{allen1999using,neely1997technical,potvin2004generating}. Such heavy work of data processing overburdens the researchers and even make some research topics impossible. Such circumstances put forward more stringent performance requirements for the infrastructure.


\subsubsection{Obstacles to apply machine learning solutions}
The financial data and task have their uniqueness and challenges. Applying the machine learning solutions to quantitative research tasks without any adaptation rarely works.  Due to the extremely low SNR(Signal to Noise Ratio) in financial data, it is very hard to build a successful data-driven strategy in financial markets. Most machine learning algorithms are data-driven and have to deal with such difficulties.
Without carefully handling the details, machine learning models can hardly achieve satisfying performance. Even a minor mistake can make the model over-fit the noise rather than learn effective patterns. Rightly handling the details requires a lot of domain knowledge of the financial industry. 
Moreover, the typical objectives, such as annualized return, are often not differentiable, which makes it hard to train models directly for machine learning methods. Defining a reasonable task with appropriate supervised targets is very important for modeling the finance data.
Such barriers daunt quite a lot of data scientists without much domain knowledge of the financial industry.

Another necessary step to build a machine learning application is hyperparameter optimization. Different machine learning algorithms have different hyperparameter search spaces, each of which has multiple dimensions with different meanings and priorities. Some of the quantitative researchers come from the traditional financial industry and don't have much knowledge about machine learning. Such huge learning cost stops many users from giving full play to the maximum value of machine learning.

\subsection{Related Work}
In the financial industry, an investment strategy will become less profitable with more investors following it. Therefore, the financial practitioners, especially quantitative researchers, are never keen to share their own algorithms and tools.
OLPS \cite{li2016olps} is the first open-source toolbox for portfolio selection. It consists of a family of classical strategies powered by machine learning algorithms as benchmarks and toolkit to facilitate the development of new learning methods. This toolbox only supports Matlab 
and Octave
, which is not compatible with current scientific mainstream language Python and thus not friendly to the modern machine learning algorithms. Its framework is quite simple, and modern quantitative research workflow based on AI technologies is much more complicated.
Other quantitative tools emerge in recent years. QuantLib \cite{firth2004use} only focuses part of modern quantitative research workflow.  QUANTAXIS \footnote{https://github.com/QUANTAXIS/QUANTAXIS} focuses more on the IT infrastructure instead of the research workflow.  Quantopian releases a series of open-source tools 1) Alphalens: a Python Library for performance analysis of predictive (alpha) stock factors 2) Zipline: an event-driven system for back-testing 3) Pyfolio: a Python library for performance and risk analysis of financial portfolios.  All of them only focus on the analysis of trading signals or an investment portfolio.

Overall, \qlibname is the first open-source platform that accommodates the workflow of a modern quantitative researcher in the age of AI. 
It aims to empower every quantitative researcher to realize the great potential of AI technologies in quantitative investment.

\section{AI-oriented Quantitative Investment Platform}

\subsection{Overall Design}
In the cooperation with the quantitative researcher with years of hands-on experience in the financial market, we've encountered all of the above problems and explored all kinds of solutions. Motivated by current circumstances, we implement \qlibname to apply AI technologies in quantitative investment.


\paragraph{AI-oriented framework} 
\qlibname is designed in a modularized way based on modern research workflow to provide the maximum flexibility to accommodate AI technologies. 
Quantitative researchers could extend the modules and build a workflow to try their ideas efficiently. In each module, \qlibname provides several default implementation choices which work very well in practical investment. With these off-the-shelf modules, quantitative researchers could focus on the problem they are interested in a specific module without distracted by other trivial details.
Besides code, computation and data can also be shared in some modules, so \qlibname is designed to serve users as a platform rather than a toolbox.


\paragraph{High-performance infrastructure} 
The performance of data processing is important to data-driven methods like AI technologies. As an AI-oriented platform, \qlibname provides a high-performance data infrastructure.
\qlibname provides a time-series flat-file database \footnote{https://en.wikipedia.org/wiki/Flat-file\_database}.  Such a database is dedicated to scientific computing on finance data. It greatly outperforms current popular storage solutions like general-purpose databases and time-series databases on some typical data processing tasks in quantitative investment research.  Furthermore, the database provides an expression engine, which could accelerate the implementation and computation of factors/features, which make research topics that rely on expressions computation possible.

\paragraph{Guidance for machine learning} \qlibname  has been integrated with some typical datasets for quantitative investment, on which typical machine learning algorithms could successfully learn patterns with generalization ability. 
\qlibname provides some basic guidance for machine learning users and integrates some reasonable tasks which consist of reasonable feature space and target label. Some typical hyperparameter optimization tools are provided. With guidance and reasonable settings, machine learning models could learn patterns with better generalization ability instead of just over-fitting the noise.
\subsection{AI-oriented Framework}
\label{modularization}

\begin{figure}[htbp]
  \includegraphics[width=\linewidth]{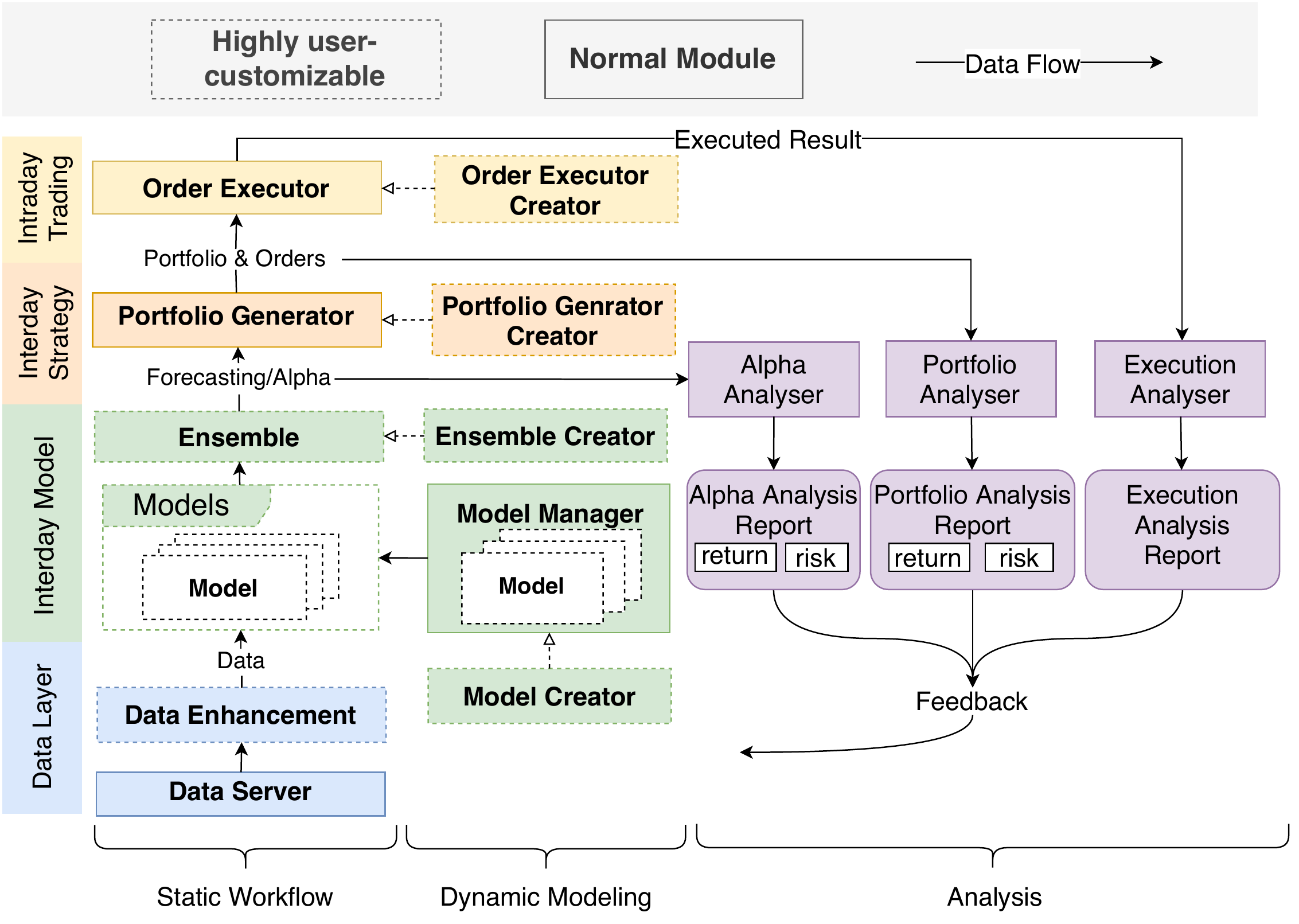}
  \caption{modules and a typical workflow built with \qlibname}
  \label{fig:framework}
\end{figure}
Figure \ref{fig:framework} shows the overall framework of \qlibnamenb. This framework aims to 1) accommodate the modern AI technology, 2) help the quantitative researchers build a whole research workflow with minimal efforts 3) and leave them the maximal flexibility to explore problems they are interested without getting distracted by other parts. 

Such a target leads to a modularized design from the perspective of system design. The system is split into several individual modules based on the modern practical research workflow. Most of the quantitative investment research directions, no matter traditional or AI-based, could be regarded as implementations of one or multiple modules' interfaces.  Qlib provides several typical implementations that work well in practical investment for users in each module . Moreover, the modules provide the flexibility for researchers to override existing methods to explore new ideas. With such a framework, researchers could try new ideas and test the overall performance with other modules with minimal cost. 

The modules of \qlibname are listed in Figure \ref{fig:framework} and connected in a typical workflow. Each module corresponds to a typical sub-task in quantitative investment. A implementation in the module can be regarded as a solution for this task. We'll introduce each module and give some related examples of existing quantitative research to show how \qlibname accommodate them. 

It starts with the \textbf{Data Server module} in the bottom left corner, which provides a data engine to  query and process raw data. With retrieved data, researcher could build his own dataset in the \textbf{Data Enhancement module}. Researchers have tried a lot solutions to build better datasets by exploring and constructing effective factors/features\cite{potvin2004generating,neely1997technical,allen1999using,kakushadze2016101}. Generating datasets for training\cite{feng2019enhancing} is another research direction to provide datasets solution. 
The \textbf{Model Creater module} learns models based on datasets. In recent years, numerous researchers have explored all kinds of models to mine trading signals from financial dataset\cite{sezer2019financial}. Moreover, meta-learning \cite{vilalta2002perspective} that tries to learn to learn provides a new learning paradigm for the Model Creator module.  Given plenty of methods to model the financial data in a modern research workflow, the model management system has become a necessary part of the workflow. The \textbf{Model Manager module} is designed to handle such problems for modern quantitative researchers. With diverse models, ensemble learning is quite an effective way to enhance the performance and robustness of machine learning models, and it is frequently used in the financial area\cite{qiu2014ensemble,yang2017stock,zhao2017deep}. It is supported by \textbf{Model Ensemble module}. \textbf{Portfolio Generator module} aims to generate a portfolio from trading signals output by models, which is known as portfolio management\cite{qian2007quantitative}. Barra \cite{sheikh1996barra} provides the most popular solution for this task. With the target portfolio, we provide a high-fidelity trading simulator, \textbf{Orders Executor module}, to examine the performance of a strategy and \textbf{Analyser modules} to automatically analyze the trading signals, portfolio and execution results.  The Order Executor module is designed as a responsive simulator rather than a back-testing function, which could provide the infrastructure for some learning paradigm(e.g., RL) that requires feedback of the environment produced by the Analyser modules. 

The data in quantitative investment are in time-series format and updated by time. The size of in-sample dataset increases by time. A typical practice to leverage the new data is to update our models regularly \cite{wang2019bi}. Besides better utilization of increasing in-sample data, dynamically updating models \cite{yang2019divide} and trading strategies \cite{wang2019conservative} will improve the performance further due to dynamic nature of the stock market\cite{adam2016stock}. Therefore, it is obviously not the optimal solution to use a set of static model and trading strategies in \textbf{Static Workflow}. Dynamic updating of models and strategies is a important research direction in quantitative investment. The modules in the \textbf{Dynamic Modeling} provide interfaces and infrastructure to accommodate such solutions.



\subsection{High Performance Infrastructure}
\label{database}

\subsubsection{Financial data}
\label{finance_data}
We'll summarise the data requirements in quantitative research in this section.
In quantitative research, the most frequently-used format of data follow such format
\begin{displaymath}
    BasicData_T = \left \{   x_{i,t,a} \right \} , i \in Inst,  t \in Time, a \in Attr
\end{displaymath}
where $x_{i,t,a}$ is the value of basic type(e.g. float, int), $Inst$ denotes the financial instruments set(e.g. stock, option, etc.), $Time$ denotes the timestampes set(e.g. trading days of stock market), $Attr$ denotes the possible attributes set of an instrument(e.g. open price, volume, market value), $T$ denote the latest timestamp of the data(e.g. the latest trading date). $x_{i,t,a}$ denotes the value of attribute $a$  of instrument $i$ at time $t$. 

Besides, instruments pools are necessary information to specify a set of financial instruments which change over time
\begin{align*}
Pool_T = \left \{ pool_t \right \}, t \in Time, pool_t \subseteq Inst
\end{align*}
S\&P 500 Index \footnote{https://en.wikipedia.org/wiki/S\%26P\_500\_Index} is a typical example of $Pool$.

Data update is an essential feature. The existing historical data will not change over time.  Only the append operation of new data is necessary.  The formalized update operation is
\begin{align*}
BasicData_T =& OldBasicData_T \cup \left \{ x_{i, t, a_{new}} \right \} \\
BasicData_{T+1} =& BasicData_T \cup \left \{ x_{i, T+1, a} \right \} \\
Pool_{T+1} =& Pool_{T} \cup \left \{  pool_{t+1} \right \}
\end{align*}
User queries can be formalized as
\begin{align*}
Data_{Query} = \{ & x_{i, t, a} | i_t \in pool_t, pool_t \in Pool_{query}  \\
                 & a \in Attr_{query}, time_{start} \leq t \leq time_{end} \}
\end{align*}
which represents data query of some attributes of instruments in a specific time range in a specific pool.


Such requirements are quite simple. Many off-the-shelf open-source solutions support such operations. We classify them into three categories and list the popular implementations in each category.
\begin{itemize}
    \item General-purpose database: MySQL\cite{mysql2001mysql}, MongoDB\cite{chodorow2013mongodb}
    \item Time-series database: InfluxDB \cite{naqvi2017time}
    \item Data file for scientific computing: Data organized by numpy\cite{oliphant2006guide} array or pandas\cite{mckinney2011pandas} dataframe
\end{itemize}

The general-purpose database supports data with diverse formats and structures.
Besides, it provides lots of sophisticated mechanisms, such as indexing, transaction, entity-relationship model, etc.  Most of them add heavy dependencies and unnecessary complexity to a specific task rather than solving the key problems in a specific scenario.  The time-series database optimizes the data structures and queries for time-series data. But they are still not designed for quantitative research, where the data are usually in compact array-based format for scientific computation to take advantage of hardware acceleration. It will save a great amount of time if the data keep the compact array-based format from the disk to the end of clients without format transformation. However, both general-purpose and time-series database store and transfer the data in a different format for the general purpose, which is inefficient for scientific computation.

Due to the inefficiency of databases, array-based data gain popularity in the scientific community.  Numpy array and pandas dataframe are the mainstream implementations in scientific computation, which are often stored as HDF\footnote{https://en.wikipedia.org/wiki/Hierarchical\_Data\_Format} or pickle\footnote{https://docs.python.org/3/library/pickle.html} on the disk.  Data in such formats have light dependencies and are very efficient for scientific computing.  However, such data are stored in a single file and hard to update or query.

After an investigation of above storage solutions, we find none could fit the quantitative research scenario very well. It is necessary to design a customized solution for quantitative research.

\subsubsection{File storage design}

\begin{figure}[htbp]
  \includegraphics[width=\linewidth]{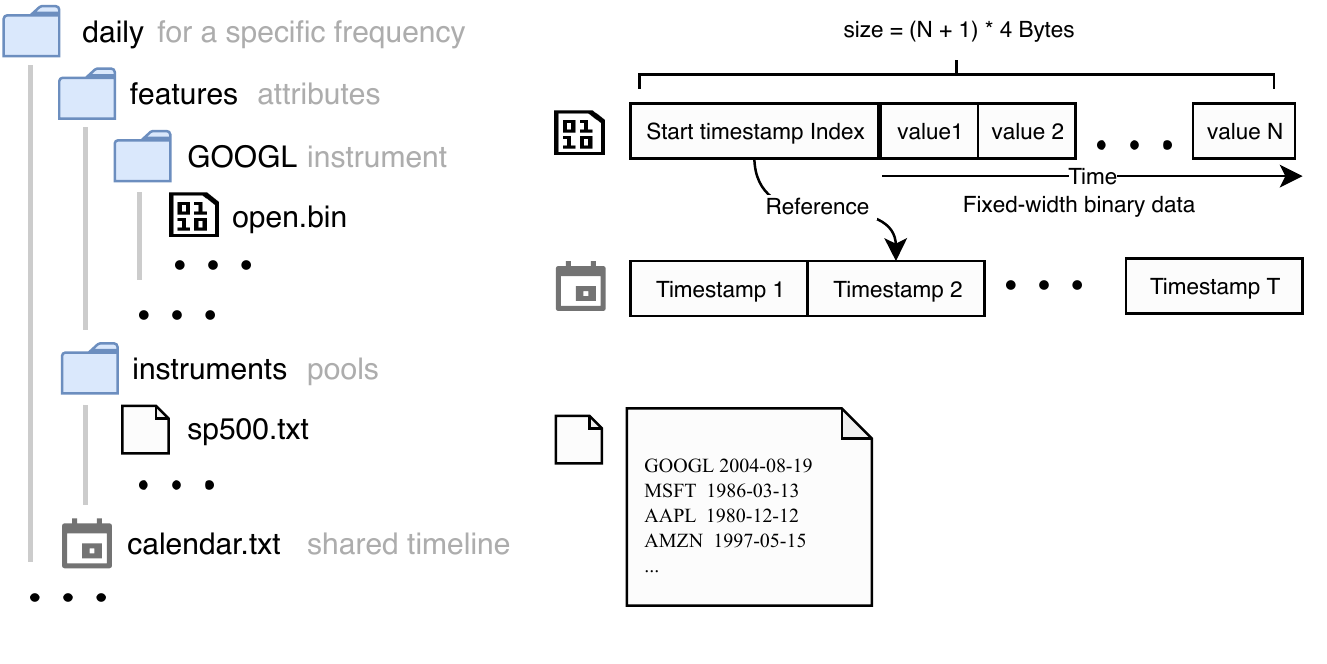}
  \caption{The description of the flat-file database; the left part is the structure of files; the right part is the content of files}
  \label{fig:file_structure}
\end{figure}

Figure  \ref{fig:file_structure} demonstrates the file storage design.
As shown in the left part of the figure, \qlibname organize files in a tree structure.
Data are separated into folders and files according to different frequencies, instruments and attributes. 
All the values of attributes are stored in binary data in a compact fixed-width format so that indexing by bytes becomes possible.  The shared timeline is stored separately in a file named "calendar.txt".  The data file of attribute values sets its first 4 bytes to the index value of the timeline to indicates the start timestamp of the series of data. With the start time index, \qlibname could align all the values on the time dimension.

The data are stored in a compact format, which is efficient to be combined into arrays for scientific computation. While it achieves high performance like array-based data in scientific computation, it meets data update requirements in the quantitative investment scenario.  All data are arranged in the order of time. New data could be updated by appending, which is quite efficient.  Adding and removing attributes or instruments are quite straightforward and efficient, because they are stored in separate files.  Such a design is extremely light-weighted. Without the overheads of databases, \qlibname achieves high performance.

\subsubsection{Expression Engine}
It is quite a common task to develop new factors/features based on basic data.  Such a task takes a large proportion of the time of many quantitative researchers.  Both Implement such factors by code, and the computation process is time-consuming. Therefore, \qlibname provides an expression engine to minimize the effort of such tasks.

Actually, the nature of factors/features is a function that transforms the basic data into the target values. The function could break down into a combination of a series of expressions. The expression engine is designed based on this idea. With this expression engine, quantitative researchers could implement new factors/features by writing expressions instead of complicated code. For example, The Bollinger Band technical indicator \cite{bollinger2002bollinger} is a widely used technical factor and its  upper bounds can be implemented by just a simple expression \emph{"(MEAN(\$close, N)+2*STD(\$close, N)-\$close)/MEAN(\$close, N)"} with the expression engine.

Such an implementation is simple, readable, reusable and maintainable. Users can easily build a dataset with just a series of simple expressions. Searching expressions to construct effective trading signals is a typical research topic, which has been explored by many researchers \cite{allen1999using,neely1997technical,potvin2004generating}. An expression engine is an essential tool for such a research topic.


\subsubsection{Cache system}

\begin{figure}[htbp]
  \includegraphics[width=\linewidth]{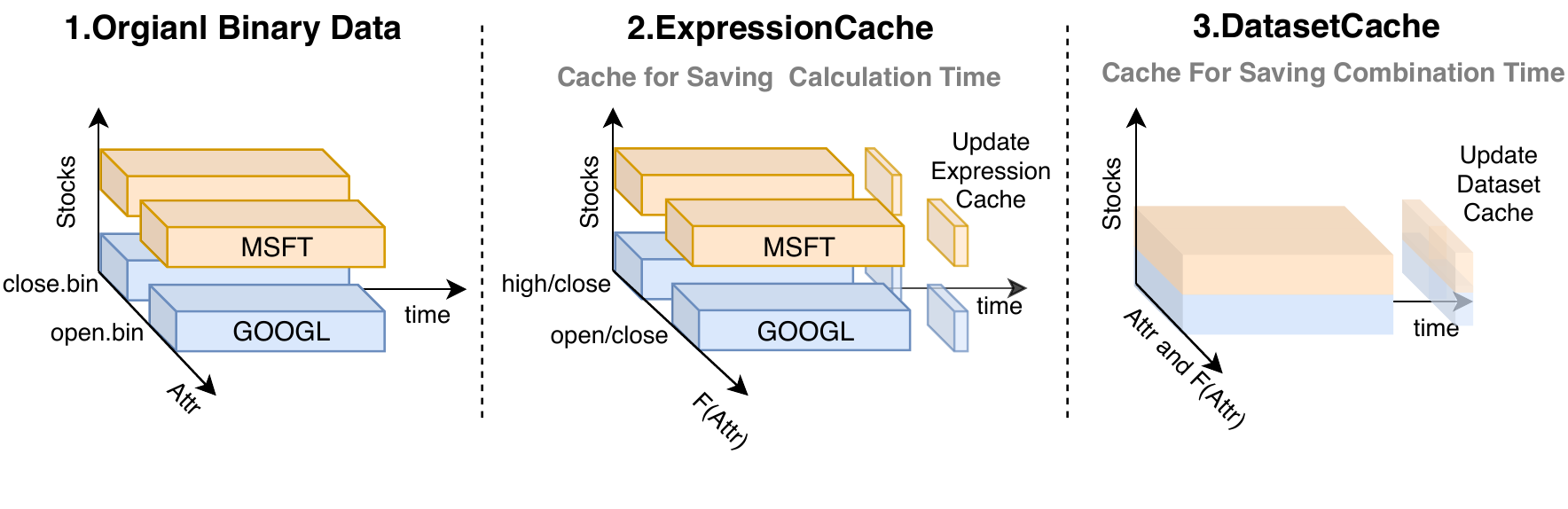}
  \caption{The disk cache system of \qlibnamenb; expression cache for saving time of expression computation; dataset cache for saving time of data combination}
  \label{fig:cache_structure}
\end{figure}

To avoid replicated computation, \qlibname has a built-in cache system.  It consists of memory cache and disk cache.

\paragraph{In-memory cache} When \qlibname computes factors/features with its expression engine, it parses the expression into a syntax tree. All computed results of nodes will be stored in an LRU(Least Recently Used) cache in memory. The replicated computation of same (sub-)expressions can be saved.

\paragraph{Disk cache} A typical workflow of data processing in quantitative investment can be divided into three steps: fetching original data, computing expressions and combining data into arrays for scientific computation. Computing expressions and combining data are very time-consuming. It could save much time if we can cache the shared intermediate data. In practical data processing tasks, many intermediate results can be shared. For example, the same expression computation can be shared by different data processing tasks.  Therefore \qlibname designed a 2-level disk cache mechanism. The cache system is shown in Figure \ref{fig:cache_structure}. The left part is the original data we described in Section \ref{finance_data}. The first level is expression cache, which will save all the computed expressions to the disk cache. The data structure of the expression cache is the same as the original data. With the expression cache, the same expression will be computed only once. After the expression cache is dataset cache, which stores the combined data to save the combination time. The cache data of both levels are arranged by time and indexable on the time dimension, so the disk cache can be shared even when the query time changes.  Moreover, \qlibname support data update by appending new data thanks to the data arrangement by time. The maintenance of the data is much easier with such a mechanism.

\begin{table*}
\centering

\begin{tabular}{l | c | c | c | c | c | c | c }  
\toprule
& HDF5 & MySQL & MongoDB &  InfluxDB  & \qlibname  -E -D & \qlibname +E -D & \qlibname +E +D \\
\midrule
\midrule
 Storage(MB) & \textbf{287} & 1,332 & 911 &  \textbf{394}  & \textbf{303} & 802 & 1,000 \\
\midrule
 Load Data(s) & \textbf{0.80$\pm$0.22}  &  182.5$\pm$4.2 & 70.3$\pm$4.9 & 186.5$\pm$1.5 & \textbf{0.95$\pm$0.05}  &  4.9$\pm$0.07  & 7.4$\pm$0.3   \\
\midrule
 Compute Expr.(s) & \multicolumn{4}{c|}{179.8$\pm$4.4 }& 137.7$\pm$7.6 & \textbf{35.3$\pm$2.3} & -   \\
\midrule
Convert Index(s) & \multicolumn{4}{c|}{-} & \multicolumn{2}{c|}{3.6$\pm$0.1} & -   \\
\midrule
Filter by Pool(s) & \multicolumn{6}{c|}{3.39 $\pm$0.24} & -   \\
\midrule
Combine data(s) & \multicolumn{6}{c|}{1.19$\pm$0.30} & -   \\

\midrule
\midrule
Total (1CPU) (s) & 184.4$\pm$3.7 & 365.3$\pm$7.5 &  253.6$\pm$6.7 &  368.2$\pm$3.6 &  147.0$\pm$8.8 & 47.6$\pm$1.0 &  \textbf{7.4$\pm$0.3}  \\
\midrule
Total(64CPUs) (s) & \multicolumn{4}{c|}{-}  &  8.8$\pm$0.6 & \textbf{4.2$\pm$0.2} &  -  \\

\bottomrule

\end{tabular}

\caption{Performance comparison of different storage solutions}
\label{tab:perf_compare}
\end{table*}

\subsection{Guidance for Machine Learning}
As we discussed in Section \ref{background}, guidance for machine learning algorithms is very important. \qlibname provides typical datasets for machine learning algorithms. Some typical task settings can be found in \qlibname, such as data pre-processing, learning targets, etc. Researchers don't have to explore everything from scratch. Such guidances provide lots of domain knowledge for researchers to start their journey in this research area. 

For most machine learning algorithms, hyperparameter optimization is a necessary step to achieve better generalization. Although it is important, it takes a lot of effort and is quite repetitive. Therefore, \qlibname provides a \textbf{Hyperparameters Tuning Engine(HTE)} to make such a task easier. 
HTE provides an interface to define a hyperparameter search space $\Theta$ and then search the best hyperparameters $\theta$ automatically. 

In a typical financial task of modeling time-series data, the new data comes in sequence by time. To leverage the new data, models have to be re-trained on new data periodically. The new best hyperparameters $\theta$ change but are often close to previous best hyperparameters. HTE provides a mechanism dedicated to hyperparameter optimization on financial tasks. It generates a new distribution for hyperparameters search space for better a chance to reach the best point with fewer trials.  The distribution for searching $\theta$ can be formalized as
\begin{equation*}
p_{new}(x) = \frac{p_{prior}(x) \varphi_{\theta_{prev},\sigma^2}\left( x \right) }{\mathbb{E}_{x \sim p_{prior}}[\varphi_{\theta_{prev},\sigma^2}\left( x \right)]}
\end{equation*}
where $p_{prior}$ is the original hyperparameters search space;  $\varphi_{\theta_{prev},\sigma^2} \left( x \right) \sim \mathcal {N}(\theta_{prev} ,\sigma ^{2})$; $\theta_{prev}$ is the best hyperparameter in last model training.   The domain of hyperparameter search space remains the same, but the probability density around $\theta_{prev}$ increases.

\section{Use Case \& Performance Evaluation}

\subsection{Use Case}
\qlibname provide a \textbf{Config-Driven Pipeline Engine(CDPE)} to help researchers build the whole research workflow show in Figure \ref{fig:framework} easier.  The user could define a workflow with just a simple config file like List \ref{fig:qlib_yaml}(some trivial details are replaced by "...").  Such an interface is not mandatory, and we leave the maximal flexibility to users to build a quantitative research workflow by code like building blocks.

\begin{figure}[htbp]
    \centering
  \includegraphics[width=0.8\linewidth]{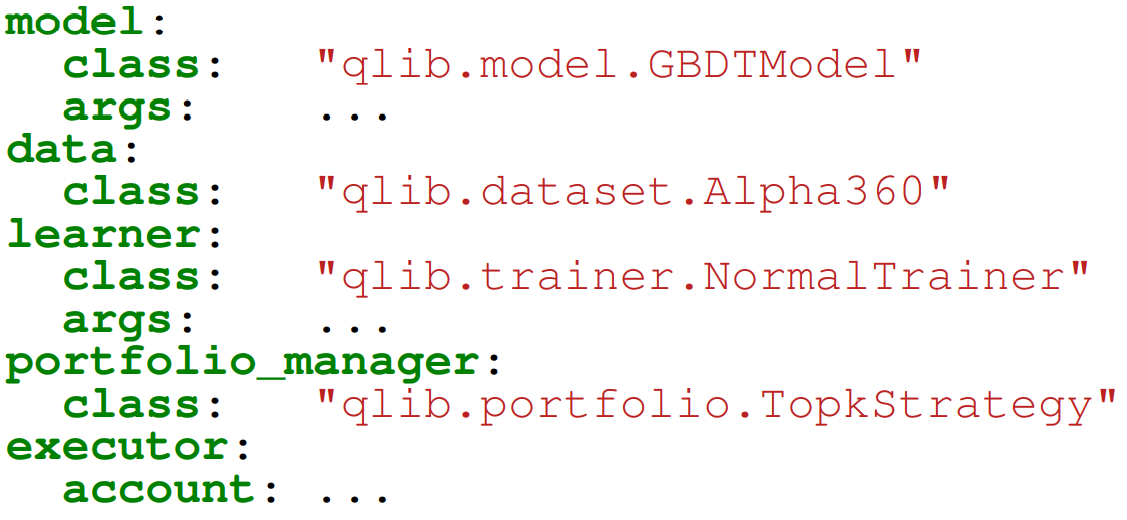}
  \caption{A Configuration example of CDPE}
  \label{fig:qlib_yam}
\end{figure}



\subsection{Performance Evaluation}

The performance of data processing is important to data-driven methods like AI technologies. As an AI-oriented platform, \qlibname provides a solution for data storage and data processing.
To demonstrate the performance of \qlibnamenb, We compare \qlibname with several other solutions
 discussed in Section \ref{database}, which includes \emph{HDF5, MySQL, MongoDB, InfluxDb and \qlibnamenb}. The \emph{\qlibname +E -D} indicates \qlibname with expression cache enabled and dataset cache disabled, and so forth.

The task for the solutions is to create a dataset from the basic OHLCV\footnote{The open, high, low, close price and trading volume of a stock} daily data of a stock market, which involves data query and processing. The final dataset consists of 14 factors/features derived from OHLCV data(e.g. "Std(\$close, 5)/\$close").  The time of the data ranges from 1/1/2007 to 1/1/2020. The stock pool consists of 800 stocks each day, which changes daily.

Besides the comparison of the total time of each solution, we break down the task into following steps for more details.

\begin{itemize}
\item \textbf{Load Data} Load the OHCLV data or cache into RAM as the array-based format for scientific computation.   
\item \textbf{Compute Expr.}  Compute the derived factors/features.
\item \textbf{Convert Index} It only applies to \qlibnamenb. Because \qlibname doesn't store the indices(i.e., timestamp, stock id) in the original data, it has to set up data indices.
\item \textbf{Filter data}  Filter the stock data by a specific pool. For example, SP500
involves more than 1 thousand stock in total, but it only includes 500 stocks daily. The data not included in SP500 on a specific day should be filtered out, though it has ever been in SP500. It is impossible to filter out data when loading data, because some derived features rely on historical OHLCV data.
\item \textbf{Combine data}  Concatenate all the data of different stocks into a single piece of array-based data
\end{itemize}

As we can seen in Table \ref{tab:perf_compare}. \qlibnamenb's compact storage achieves similar size and loading speed as the dedicated scientific HDF5 data file. The databases take too much time on loading data.  After looking into the underlying implementation, we find that data go through too many layers of interfaces and unnecessary format transformations in both general-purpose database and time-series database solution. Such overheads greatly slow down the data loading process. Due to the memory cache of \qlibnamenb,  \qlibname -E -D saves about 24\% of the time of Compute Expr. Moreover, \qlibname provides expression cache and dataset cache mechanism.
With expression cache enabled in \qlibname +E -D,  80.4\% of the time for Compute Expr. is saved if no expression cache is missed. Combining the factors/features into one piece of array-based data for each stock accounts for the major time consuming of \qlibname +E -D, which is included in the Compute Expr. step. Besides the computation cost,  the most time-consuming step is data combination. The dataset cache is designed to reduce such overheads.  As shown in the column \qlibname +E +D,  the time cost is further reduced.

Moreover, \qlibname can leverage multiple CPU cores to accelerate computation. As we can see in the last line of Tabel \ref{tab:perf_compare}, the time cost is significantly reduced for \qlibname with multiple CPUs. \qlibname +E +D can't be accelerated further due to it just reads the existing cache and almost computes nothing.

\subsection{More about Qlib}
Qlib an opensource platform in continuous development. More detailed documentations can be found in its github repository \footnote{https://github.com/microsoft/qlib/}. A lot of features(e.g. data service with client-server architecture, analysis system, automatic deployment on the cloud) not introduced in detail in this paper could be found in the online repository. Your contributions are welcomed.

\section{Conclusion} 
In this paper, we present practical problems of modern quantitative researchers in the age of AI. Based on these practical problems, we design and implement \qlibname that aims to empower every quantitative researcher to realize the great potential of AI-technologies in quantitative investment.


\bibliographystyle{named}
\bibliography{main}

\end{document}